\documentclass[slac_one]{revtex4}
\usepackage{graphicx}
\usepackage{fancyhdr}
\begin{document}

\newcommand{\dedx}{\mbox{${\rm d}E/{\rm d}x$}}

\title{A critical appraisal of the LSND anomaly} 

%

\author{I. Boyko (for the 
HARP--CDP group\footnote{The members of the 
HARP--CDP group are:
A.~Bolshakova, 
I.~Boyko, 
G.~Chelkov, 
D.~Dedovitch, 
A.~Elagin, 
M.~Gostkin,
S.~Grishin,
A.~Guskov, 
Z.~Kroumchtein, 
Yu.~Nefedov, 
K.~Nikolaev and
A.~Zhemchugov 
from the Joint Institute for Nuclear Research, 
Dubna, Russian Federation;
F.~Dydak and  
J.~Wotschack 
from CERN, Geneva, Switzerland; 
A.~De~Min 
from the Politecnico di Milano and INFN, 
Sezione di Milano-Bicocca, Milan, Italy; and
V.~Ammosov, 
V.~Gapienko, 
V.~Koreshev, 
A.~Semak, 
Yu.~Sviridov, 
E.~Usenko and  
V.~Zaets 
from the Institute of High Energy Physics, Protvino, 
Russian Federation.} 
)}
\affiliation{Joint Institute for Nuclear Research, Dubna,
Russian Federation}

\begin{abstract}
The so-called `LSND anomaly', a 3.8$\sigma$ excess 
of $\bar{\nu}_{\rm e}$ events interpreted as originating 
from $\bar{\nu}_{\mu} \rightarrow \bar{\nu}_{\rm e}$
oscillation, gave rise to many theoretical speculations.
The MiniBooNE Collaboration reported inconsistency of this 
interpretation with the findings from their search for
$\nu_{\mu} \rightarrow \nu_{\rm e}$ oscillations.   
Yet the origin of the LSND anomaly was never clarified.
A critical issue is the prediction of the 
background $\bar{\nu}_{\rm e}$ flux that was used in the 
analysis of the LSND experiment.
For this, decisive input comes from pion spectra measured with 
the HARP large-angle spectrometer under conditions that closely
resemble the LSND situation: a proton beam with 800~MeV kinetic
energy hitting a water target.
\end{abstract}

\maketitle

\thispagestyle{fancy}


\section{INTRODUCTION} 

The LSND experiment at Los Alamos studied the
$\bar{\nu}_{\rm e}$ flux originating from protons with
800~MeV kinetic energy hitting a beam dump consisting  
of water to a good fraction. They reported 
a 3.8$\sigma$ excess of $\bar{\nu}_{\rm e}$ events 
over background~\cite{LSNDexperiment}. The excess was interpreted 
as originating from $\bar{\nu}_{\mu} \rightarrow \bar{\nu}_{\rm e}$
oscillation, which gave rise to many theoretical speculations
as to the existence of sterile neutrinos. 

The MiniBooNE Collaboration reported inconsistency of this 
interpretation with the findings from their search at FNAL
for $\nu_{\mu} \rightarrow \nu_{\rm e}$ 
oscillations~\cite{MiniBooNE}. 

A critical issue in the LSND analysis is the background level 
of $\bar{\nu}_{\rm e}$ events that originates from the decay chain
$\pi^- \rightarrow \mu^- \rightarrow \bar{\nu}_{\rm e}$. 
An underestimate of $\pi^-$ production which was quite
uncertain at the time of the LSND experiment, would reduce the
anomalous excess of $\bar{\nu}_{\rm e}$ events. The
relevant input is a precise measurement of the ratio of 
secondary $\pi^-$ to $\pi^+$ production by 800~MeV protons.

With a view to clarifying this issue, the HARP detector at the 
CERN PS that took data in 2001 and 2002
with proton and pion beams with momentum from 
1.5 to 15~GeV/{\it c}, also recorded data from the exposure 
of a water target to protons with 800~MeV kinetic 
energy (1.5~GeV/{\it c} momentum).

This paper reports ratios $\pi^-/\pi^+$ from this exposure and
compares them with the ratios that were used in the LSND
analysis.

\section{DETECTOR CHARACTERISTICS AND PERFORMANCE}

\begin{figure*}[h]
\begin{center}
\begin{tabular}{cc} 
\includegraphics[height=0.3\textwidth]{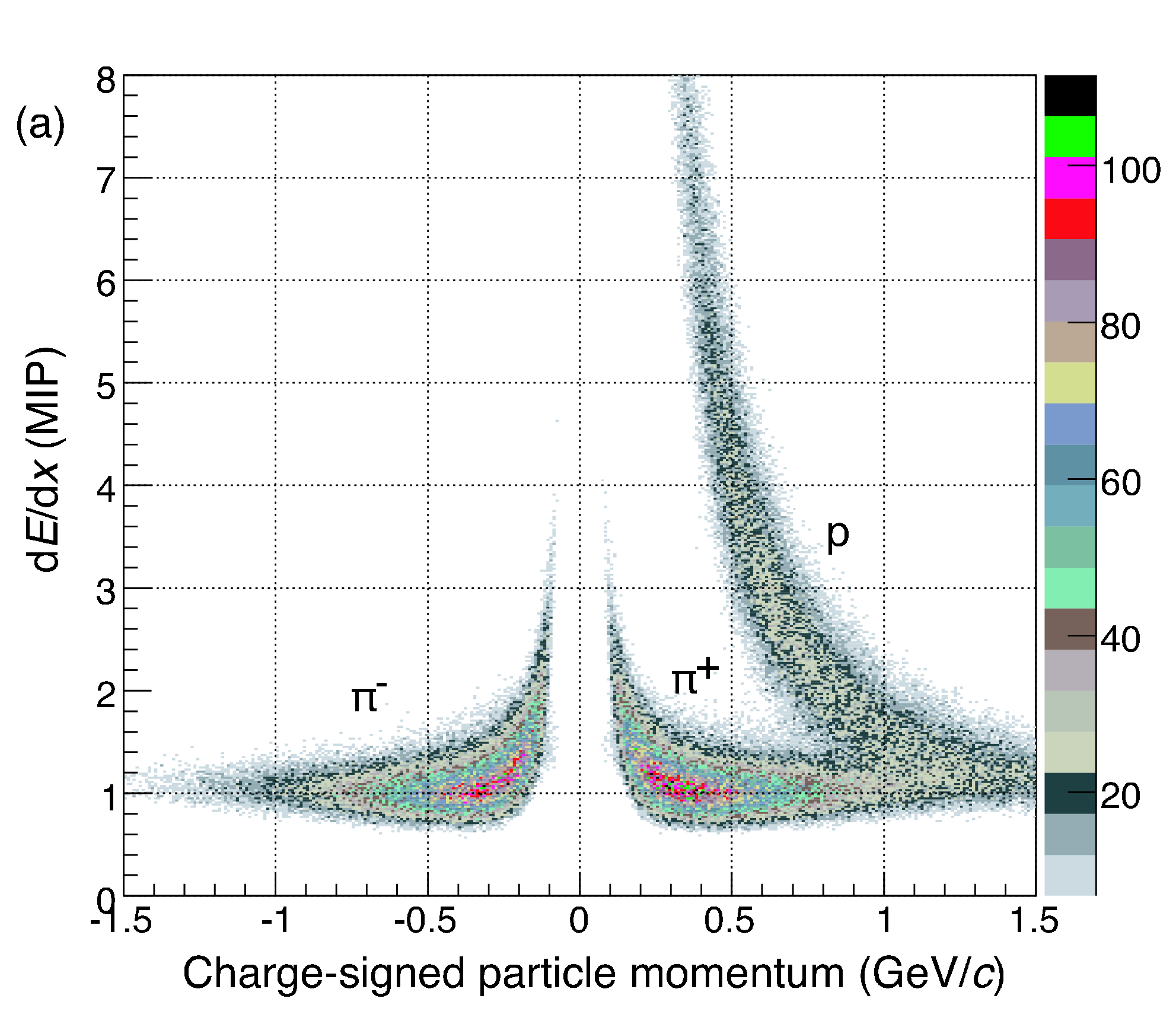} &
\includegraphics[height=0.3\textwidth]{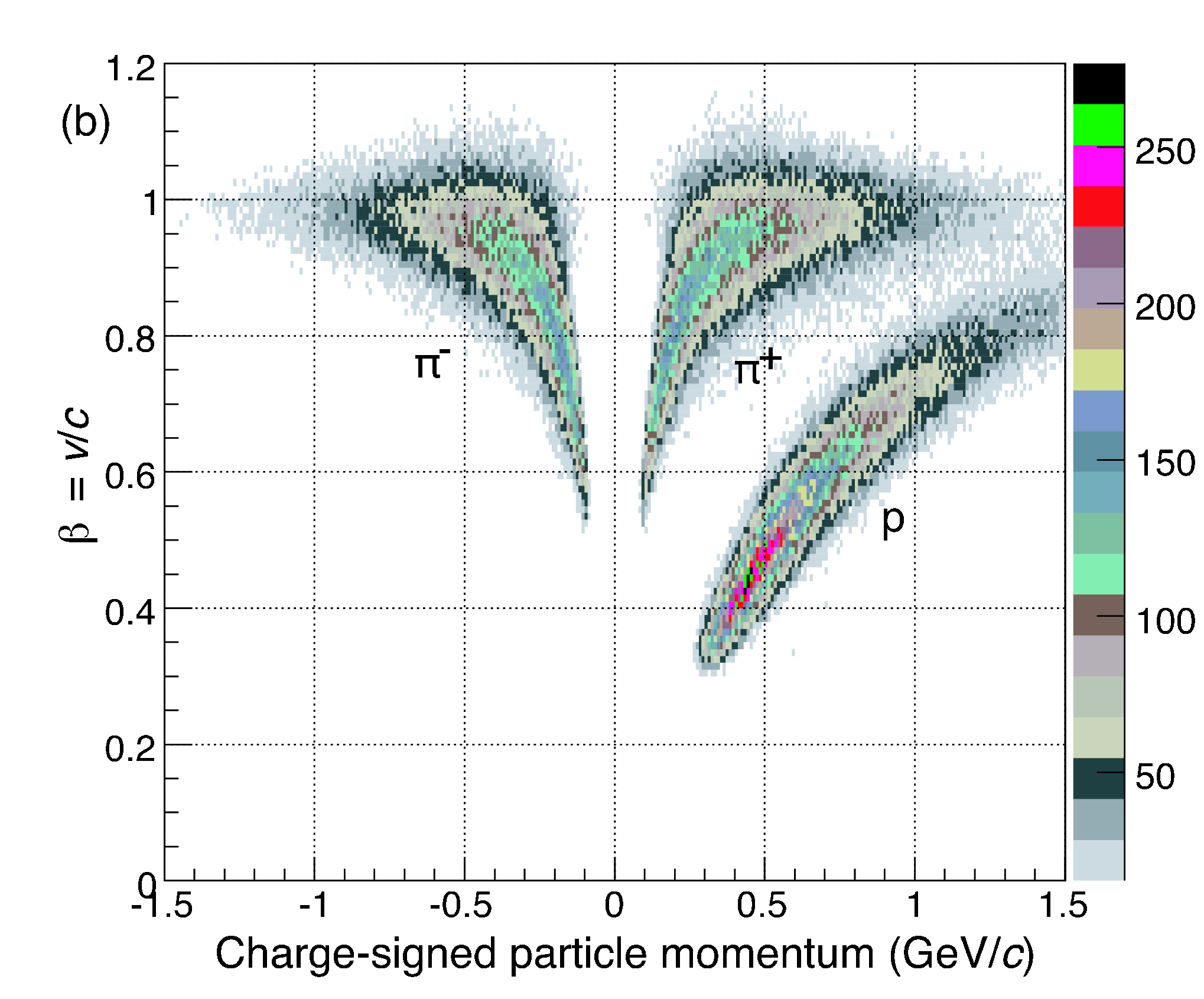} \\
\end{tabular}
\end{center}
\caption{Specific ionization d$E$/d$x$ (left panel) and velocity $\beta$
(right panel) versus the charge-signed momentum of positive and 
negative tracks 
in $+8.9$~GeV/{\it c} data.}
\label{dedxandbeta}
\end{figure*}

\begin{figure*}[h]
\begin{center}
\includegraphics[width=0.55\textwidth]{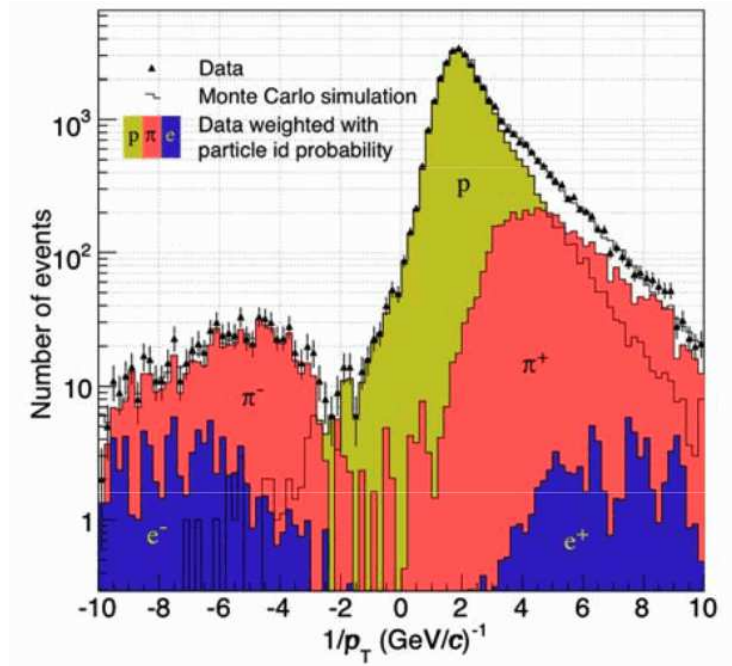}
\end{center}
\caption{Decomposition, on a logarithmic scale, of the observed 
spectrum of secondaries from
the interaction of 800~MeV protons in water into particle species,
as a function of the charge-signed $1/p_{\rm T}$.}
\label{particleid}
\end{figure*}

The HARP detector combined a forward spectrometer with a 
large-angle spectrometer. The latter comprised a 
cylindrical Time Projection 
Chamber (TPC) around the target and an array of 
Resistive Plate Chambers (RPCs) that surrounded the 
TPC. The purpose of the TPC was track 
reconstruction and particle identification by \dedx . The 
purpose of the RPCs was to complement the 
particle identification by time of flight.
For the work reported here, only the HARP large-angle spectrometer
was used~\cite{TPCpub,RPCpub}. Its salient technical 
characteristics are stated in Table~\ref{LAcharacteristics}. The 
good particle identification 
capability stemming from \dedx\ in the TPC and from time of flight 
in the RPC's is demonstrated in Fig.~\ref{dedxandbeta}. Correct
particle identification is of crucial importance
as 800~MeV protons produce many more secondary 
protons than $\pi^+$'s. This is highlighted in Fig.~\ref{particleid}
which demonstrates that the decomposition of the observed secondary 
particle spectrum into particle species is well understood.
\begin{table}[ht]
\vspace*{2mm}
\begin{center}
\caption{Technical characteristics of the HARP large-angle spectrometer}
\begin{tabular}{|c|c|}
\hline 
\textbf{TPC} & \textbf{RPCs}  \\
\hline
\hline
$\sigma(1/p_{\rm T}) \sim 0.20-0.25$~(GeV/{\it c})$^{-1}$ &
   Intrinsic efficiency $\sim 98$\%  \\
$\sigma(\theta) \sim 9$~mrad  & $\sigma$(TOF) $\sim 175$~ps  \\
$\sigma({\rm d}E/{\rm d}x) / {\rm d}E{\rm d}x \sim 0.16$  &  \\ 
\hline
\end{tabular}
\label{LAcharacteristics}
\end{center}
\end{table}

\clearpage

\section{HARP--CDP MEASUREMENTS VERSUS LSND MONTE CARLO SIMULATIONS}

Most positive pions from the interaction of the 800~MeV protons in the
beam dump are slowed down and decay at rest, while a few per cent decay
at flight. The positive muons from the decay also slow down and
mostly decay at rest, with only a very small fraction decaying in 
flight. The neutrino
flux resulting from positive pions consists of $\nu_\mu$, $\nu_{\rm e}$, and
$\bar{\nu}_\mu$. For negative pions the decay chain is
the same except for charge conjugation, yet negative pions that come
to rest disappear 100\% by strong interaction, and likewise negative 
muons at the 90\% level by weak interaction. The neutrino flux 
resulting from negative pions can only come from pions decaying in
flight, and muons decaying in orbit after capture. The neutrino 
flux consists of $\bar{\nu}_\mu$, $\bar{\nu}_{\rm e}$, and
$\nu_\mu$. Overall, the $\bar{\nu}_{\rm e}$ which is of interest is reduced
by a factor of order $10^{-4}$. In more detail, the reduction depends
on the level of $\pi^-$ production, the probability of $\pi^-$ decay 
in flight (related to the momentum spectrum and the beam dump geometry), 
and of the probability
of a stopped $\mu^-$ to decay in orbit.

\begin{figure*}[h]
\vspace*{3mm}
\begin{center}
\includegraphics[width=0.6\textwidth]{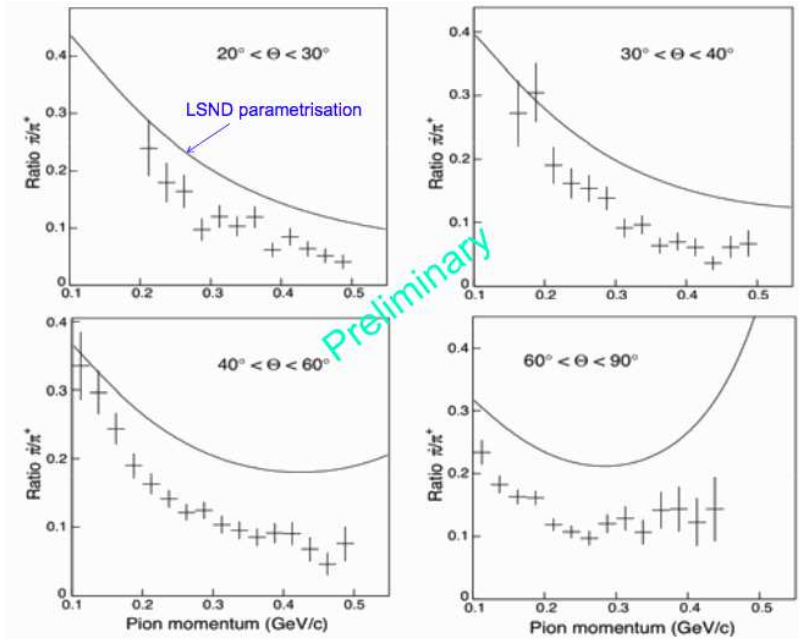} \\
\end{center}
\caption{Ratios $\pi^-/\pi^+$ in four bins of polar angle $\theta$
with respect to the incoming proton direction, as a function of
pion momentum; the crosses denote data, the lines 
represent the parametrization that was used by LSND.}
\label{pionratio}
\end{figure*}

Our measurement addresses the first of these three issues. 
Figure~\ref{pionratio} shows preliminary results of the 
measured ratio of $\pi^-$ to $\pi^+$
production in four bins of polar angle $\theta$
with respect to the incoming proton direction. Also shown is the
parametrization of this ratio that was used in the 
LSND analysis~\cite{Sung}. The measured ratio
is smaller than the parametrization. Thus, if by the time
of the LSND analysis our data had been known, the reported anomalous
excess of $\bar{\nu}_{\rm e}$ events would have been even larger.
Further work on this interesting result is in progress.


\vspace*{-8mm}

\begin{thebibliography}{9}   

\bibitem{LSNDexperiment} A.~Aguilar {\it et al.}, 
Phys. Rev. {\bf D64} (2001) 112007

\bibitem{MiniBooNE} A.A.~Aguilar--Arevalo {\it et al.},
Phys. Rev. Lett. {\bf 98} (2007) 231801 

\bibitem{TPCpub} V.~Ammosov {\it et al.},
Nucl. Instrum. Methods Phys. Res. {\bf A588} (2008) 294

\bibitem{RPCpub} V.~Ammosov {\it et al.},
Nucl. Instrum. Methods Phys. Res. {\bf A578} (2007) 119  

\bibitem{Sung} We thank Myungkee Sung for providing the 
LSND parametrization; the program code is available at
{\it http://hep.phys.lsu.edu/sung/lsnd/beammc/index.html}

\end{thebibliography}
\end{document}